\documentclass[prl,aps,draft,showpacs]{revtex4}
\usepackage[dvips,final]{graphicx}
\usepackage{bm}
\usepackage{amssymb}

\begin{document}

\title{Coarsening Dynamics of Biaxial Nematic Liquid Crystals}
\author{N.~V. Priezjev}
\author{ Robert A. Pelcovits}
\affiliation{Department of Physics, Brown University, Providence, RI 02912}
%\date{\today}
\begin{abstract}
We study the coarsening dynamics of two and three dimensional biaxial nematic liquid crystals, using Langevin dynamics. Unlike previous work, we use a model with no a priori relationship among the three elastic constants associated with director deformations. We find a rich variety of coarsening behavior, including the simulataneous decay of nearly equal populations of the three classes of half--integer disclination lines. The behavior we observed can be understood on the basis of the relative values of the elastic constants and the resulting decay channels of the defects. 

 \end{abstract}
\pacs{64.70Md, 61.30Jf}
\maketitle

Topological defects play an important role in the equilibration process following a quench from a disordered to an ordered phase (``coarsening dynamics"). 
Coarsening dynamics in nematic liquid crystals, particularly with uniaxial ordering, has been the subject of much active investigation in recent years in theory, experiments and simulations \cite{Blundell:92,Zapotocky:95,Chuang:93,Billeter:99}, in part because of the rich defect structure of liquid crystals.
On the other hand, relatively little attention (with the exception of the two--dimensional work of Ref. \cite{Zapotocky:95}) has been paid to coarsening dynamics in biaxial liquid crystals, in part because of the dearth of experimental realizations of biaxial liquid crystalline phases. However, biaxial nematics have many unusual topological features, which might be expected to influence their coarsening dynamics and thus warrant study. Biaxial nematics differ from their uniaxial counterparts in that they possess four topologically distinct classes of line defects (disclinations), while possessing no stable point defects (except in two dimensions where the line defects reduce to points) \cite{Toulouse:77,Mermin:79}. The classes of disclination lines are distinguished by the rotation of the long and short axes of the rectangular building blocks of the system. In the first three classes one of the three axes is uniformly ordered, while the remaining two axes rotate by $180^\circ$ about the core of the defect. The fourth class consists of $360^\circ$ rotations of two of the three axes. The disclination lines form closed loops in three dimensions (with a single defect class per loop) or form a network where three lines, each from a different class meet at junction points \cite{Deneve:92}. The fundamental homotopy group of biaxial nematics is non--Abelian leading to a number of interesting consequences. E.g., the merging of two defects will depend on the path they follow, and two $180^\circ$ disclinations of different types will be connected by a $360^\circ$ ``umbilical" cord after crossing each other \cite{Poenaru:77}. 
   
Zapotocky et al.~\cite{Zapotocky:95} studied coarsening dynamics in a two--dimensional model of biaxial nematics, utilizing a cell--dynamical scheme applied to a Landau--Ginzburg model, where the gradient portion of the energy was given by:
\begin{equation}\label{zap}
F_{\text{grad}}=\frac{1}{2}M (\partial_{\alpha}Q_{\beta\gamma})(\partial_{\alpha}Q_{\beta\gamma}).
\end{equation}
Here $M$ is a coupling constant, and $Q_{\alpha \beta}$ is the symmetric--traceless nematic order parameter tensor. Repeated indices are summed over; in the case of a two--dimensional nematic, $\alpha$ is summed over $x$ and $y$, while $\beta$ and $\gamma$ are summed over $x, y$ and $z$. Zapotocky et al. found that of the four topologically distinct classes of disclinations, only two classes (both corresponding to ``half-integer" lines, i.e., $180^\circ$ rotations) were present in large numbers at late times. Subsequently, Kobdaj and Thomas \cite{Kobdaj:94} showed within this one--elastic constant approximation that one class of half--integer disclinations is always energetically unstable towards dissociation into disclinations of the other two half--integer classes.

In this Letter we show that if one considers a more general gradient energy the coarsening dynamics of biaxial nematics is much richer than what occurs with the above simple model.  In particular, with appropriate sets of parameters one can obtain a coarsening sequence with all three classes of half--integer disclinations present in nearly equal numbers even at late times, or a sequence with only one class of half--integer disclinations surviving until late times. When all three classes are present, the topology of the coarsening sequence in three dimensions is markedly different from the uniaxial case.
 
To understand why the model free energy of Eq. (\ref{zap}) is not general enough for coarsening studies, it is helpful to see what it yields for the director elastic constants. For biaxial nematics there are three directors which form an orthonormal triad of vectors $\mathbf{u,v,w}$  describing the alignment of the constituent ``brick--like" molecules. The tensor $Q_{\alpha \beta}$ can be written in terms of the orthonormal triad  as:
\begin{equation}\label{Q}
Q_{\alpha \beta} =S(w_{\alpha}w_{\beta}-\frac{1}{3}\delta_{\alpha \beta})+T(u_{\alpha}u_{\beta}-v_{\alpha}v_{\beta}),
\end{equation}
where $S$ and $T$ are respectively the uniaxial and biaxial order parameters. If we insert Eq. (\ref{Q}) into Eq. (\ref{zap}), we find \cite{Sukumaran:97}:
\begin{widetext}
\begin{eqnarray}\label{general}
F_{\text{grad}}&=&K_u\lbrack(\mathbf{u \cdot \nabla v \cdot w})^2+(\mathbf{v\cdot\nabla v \cdot w})^2+(\mathbf{w \cdot \nabla w \cdot v})^2\rbrack +K_v\lbrack(\mathbf{v \cdot \nabla w \cdot u})^2+(\mathbf{u\cdot\nabla u \cdot w})^2+(\mathbf{w \cdot \nabla w \cdot u})^2\rbrack\nonumber\\&+&K_w\lbrack(\mathbf{w \cdot \nabla u \cdot v})^2+(\mathbf{u\cdot\nabla u \cdot v})^2+(\mathbf{v \cdot \nabla v \cdot u})^2\rbrack,
\end{eqnarray}
\end{widetext}
where
\begin{eqnarray}\label{K}
K_u&=&2 M (S+T)^2\nonumber\\
K_v&=&2 M (S-T)^2\nonumber\\
K_w&=&8 M T^2
\end{eqnarray}
Thus, there are three elastic constants $K_u,K_v,K_w$, each corresponding to one of the three classes of line defects. Each class corresponds to rotations of two of the three vector fields $\mathbf{u,v,w}$, about the defect core, with the third vector of the triad undistorted. We denote the classes as follows \cite{Mermin:79,Deneve:92}: $C_{u}$ ($\mathbf u$ undistorted), $C_{w}$ ($\mathbf w$ undistorted), and $C_{v}$ ($\mathbf v$ undistorted). The energy of a defect in class $C_{i}$, $i=u,v,w$ is proportional to the elastic constant $K_i$.
 Note, however, that in the model specified by Eq. (\ref{zap}) only two of the three elastic constants are independent, as they are related as indicated in Eq. (\ref{K}). In fact, the specific relationship among the three elastic constants gives rise, as shown in Ref. \cite{Kobdaj:94}, to the presence of only two defect classes at late times. Irrespective of the values of $S$ and $T$ ($M$ simply sets the overall scale of all three elastic constants), one of the three elastic constants is \textit{always} greater than the other two, yielding a decay channel for the defect in the class with the largest elastic constant.

There is no symmetry reason to restrict our attention to the model free energy, Eq.~(\ref{zap}). Even if we neglect the elastic anisotropy associated with bend, splay and twist distortions, a biaxial nematic should be described in general by three \textit{independent} elastic constants, $K_u,K_v$ and $K_w$ \cite{Sukumaran:97}. In terms of the order parameter tensor $Q_{\alpha \beta}$, this requires a term of third order in $F_{\text{grad}}$:
\begin{equation}
F_{\text{grad}}=\frac{1}{2}M (\partial_{\alpha}Q_{\beta\gamma})(\partial_{\alpha}Q_{\beta\gamma})+L Q_{\alpha \beta}(\partial_{\rho}Q_{\alpha \gamma})(\partial_{\rho}Q_{\beta \gamma}),
\end{equation}
which upon substituting Eq.~(\ref{Q}) yields the elastic constants:
\begin{eqnarray}\label{Kgeneral}
K_u&=&2 (M (S+T)+LST)(S+T)\nonumber\\
K_v&=&2 (M (S-T)-LST)(S-T)\nonumber\\
K_w&=&4(2 M-LT) T^2
\end{eqnarray}
Because of the extra coupling constant $L$, there is no predetermined hierarchy among these elastic constants. Unlike the model of Eq. (\ref{zap}) it is now possible with appropriate choices of $S, T$ and the ratio $L/M$ to have all three elastic constants equal (which will give rise as we demonstrate below to a coarsening sequence with nearly equal populations of the three classes of half--integer defects), or have one constant smaller than the remaining two, yielding a coarsening sequence with only one class of defects at late times, or recover the behavior seen in Ref. \cite{Zapotocky:95}. 

To simulate the coarsening dynamics of this general model of biaxial nematics, we consider its lattice analog introduced by Straley \cite{Straley:74}.  In this model the interaction between two biaxial objects located at sites \textit{i} and \textit{j} of a cubic lattice with orientations specified by the orthonormal triad $\mathbf{u,v,w}$ is given by:
\begin{eqnarray}
\label{Uij}
 U_{ij} = -\frac{3}{2}\beta (\mathbf w_{i} \cdot\mathbf  w_{j})^2  -
2\gamma [ (\mathbf u_{i} \cdot \mathbf u_{j})^2 - 
(\mathbf v_{i} \cdot \mathbf v_{j})^2 ] - \nonumber\\
\frac{\delta}{2} [ (\mathbf u_{i} \cdot \mathbf u_{j})^2 + 
(\mathbf v_{i} \cdot \mathbf v_{j})^2  -
\ (\mathbf u_{i} \cdot \mathbf v_{j})^2 -
(\mathbf v_{i} \cdot \mathbf u_{j})^2 ]. 
\end{eqnarray}
This model has a phase diagram with two uniaxial phases, one with rodlike order (alignment of the $\mathbf{w}$ vector field), one with discotic order (alignment of the $\mathbf{u}$ vector field) and a biaxial phase with alignment of all three vector fields \cite{Straley:74,Luckhurst:80,Biscarini:95}. 

The elastic constants emerging from Eq. (\ref{Uij}) can be determined by considering the interaction between two objects which are aligned in turn along each of the three directions $\mathbf{u,v,w}$, with the results:
\begin{eqnarray}\label{KStraley}
K_u&=&\frac{3\beta}{2}-2 \gamma +\frac{\delta}{2}\nonumber\\
K_v&=&\frac{3\beta}{2}+2 \gamma +\frac{\delta}{2}\nonumber\\
K_w&=&2 \delta
\end{eqnarray}
As in the continuum model Eq.~(\ref{general}), the parameters $\beta,\gamma$ and $\delta$ give rise to three independent elastic constants. Stability requires that $3\beta + \delta>4 |\gamma|$, and $\beta, \delta >0$.

We have simulated the coarsening dynamics associated with Eq.~(\ref{Uij}) using Langevin dynamics,
expressing the three unit vectors $\mathbf w$, $\mathbf u$ and 
$\mathbf v$ in terms of Euler angles $\phi$, $\theta$ and $\psi$. 
The equations of motion are given by :
\begin{eqnarray}\label{langevin}
\zeta \frac{\partial\phi_i}{\partial t} &=& - \frac{\partial U}{\partial \phi_i} + R_{\phi}(t)\nonumber\\
\zeta \frac{\partial\cos\theta_i}{\partial t} &=& - \frac{\partial U}{\partial \cos\theta_i} + R_{\theta}(t)\nonumber\\
\zeta \frac{\partial\psi_i}{\partial t}& =& - \frac{\partial U}{\partial \psi_i} + R_{\psi}(t)
\end{eqnarray}
where $U$ is the sum of $U_{ij}$ over the nearest--neighbors of site \textit{i}, $\zeta$ is a damping coefficient and $R_{\phi}(t),R_{\theta}(t)$ and $R_{\psi}(t)$ are uncorrelated random thermal noise sources. Each of the random variables $R$ has a Gaussian distribution of variance $2 k_B T \zeta/\delta t$ where $k_{B}$, $T$ and $\delta t$ are Boltzmann's constant, the temperature and
the time step respectively. We measured time in units of $\beta$ (choosing $\zeta=1$), and used a timestep $\delta t = 0.0005$. A dynamical equation for $\rm{cos} \  \theta$ rather than $\theta$ must be used in order to reach the correct equilibrium states \cite{Bac:01}. We verified that our dynamical equations led to the same phase diagram produced by Monte Carlo simulations \cite{Luckhurst:80,Biscarini:95}.

As in previous numerical studies of defect behavior \cite{Lammert:95,Priezjev:01}, we introduce 
a disclination line segment counting operator,
\begin{equation}
\label{D}
D^w_{ijkl} \equiv \frac{1}{2}\lbrack{1-{\rm sgn}\{(\mathbf w}_{i} \cdot
{\mathbf w}_{j})({\mathbf w}_{j} \cdot {\mathbf w}_{k})({\mathbf w}_{k} \cdot 
{\mathbf w}_{l})({\mathbf w}_{l} \cdot {\mathbf w}_{i}) \} \rbrack
\end{equation}
which is unity if a disclination line segment pierces the lattice square defined
by the four vectors ${\mathbf w}_i,{\mathbf w}_j,{\mathbf w}_k$ and ${\mathbf w}_l$. 
We define analogous operators $D^u_{ijkl}$ and $D^v_{ijkl}$ for the
$\mathbf{u}$ and $\mathbf{v}$ vectors on this lattice square. In
principle either two or none of the three operators should be unity
for a given square, and thus we can assign the line segment to one of
the classes $C_{u}$, $C_{v}$ or $C_{w}$. In practice, we found a small
number of squares where only one or all three operators were unity, an artifact of the discreteness of the underlying lattice. We obtained physically reasonable results by classifying the defects on the basis of the operators $D^u_{ijkl}$ and $D^w_{ijkl}$, assuming that $D^v_{ijkl}$ was unity only if \textit{one} of the two former operators was unity. This procedure always yielded closed defect loops in three dimensions, a reasonable test of our algorithm. We determined the location of the integer--valued defects where either $\mathbf w$ or $\mathbf u$ (or both) rotate by $\pm 360 ^{\circ}$ degrees using the algorithm of Ref. \cite{Billeter:99} which provides an upper bound on these defects. We found very few integer--valued defects using this method, so a more accurate algorithm is not needed. 

 We quenched the system instantaneously from an initial configuration of random orientations of the three vectors $\mathbf{u,v,w}$ (i.e., a high temperature state) to zero temperature (a biaxially ordered state for all $\delta \not= 0$), and then let the system evolve in time according to the dynamical equations (\ref{langevin}) monitoring the defect populations both visually and statistically. We studied the model in both two and three dimensions.  

\begin{figure}
\includegraphics[scale=0.5,angle=270]{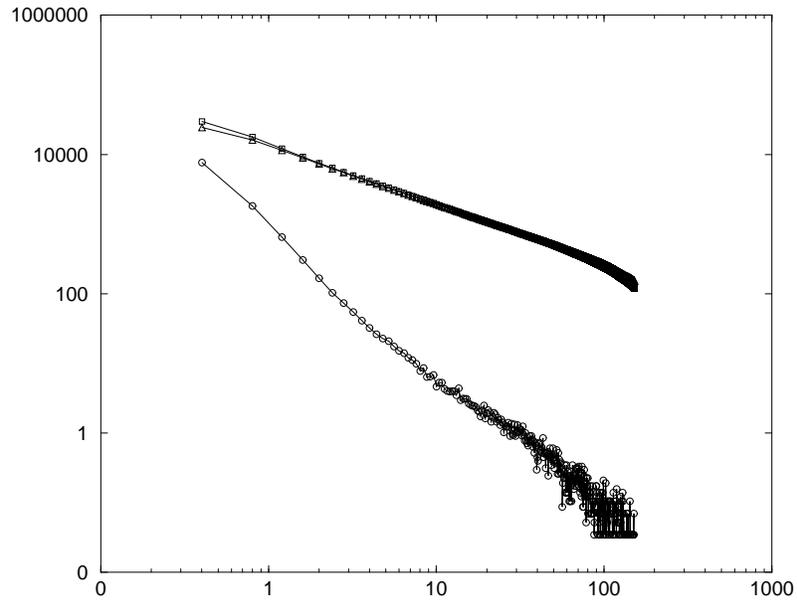}
\caption{Log-log plot of the total line length of $C_{v} \ (\circ), 
C_{w} \ (\rm{box})$, and $ C_{u} \ (\triangle)$ class defects for \ system size $40^{3}$, and parameters $\gamma=1/2$, $\delta = 4 \gamma^{2} /3= 1/3$,  as a function of time after the
quench. The data has been averaged over 60 initial random configurations.}
\label{fig1}
\end{figure}

\begin{figure}
\includegraphics[scale=0.5]{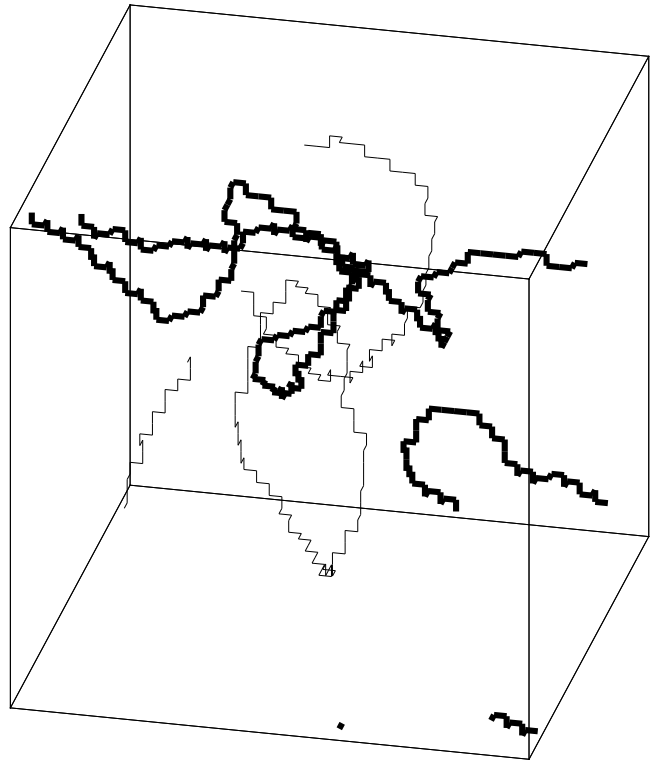}

\caption{The configuration of $C_{u}$ (thin line) and $C_{w}$ (bold line) class 
 defects for system size $40^{3}$, and parameters  $\gamma=1/2$, $\delta = 4 \gamma^{2} /3= 1/3$, at dimensionless time 80 after the quench. 
Note that all $C_{v}$ class defects have vanished at this late time, therefore $C_{w}$ and $C_{u}$ form nonintersecting loops which coarsen independently.}
\label{fig2}
\end{figure}

In agreement with our arguments above regarding the elastic constants,
we found three qualitatively distinct types of coarsening behavior
depending on the values of $\gamma$ and $\delta$ (we set
$\beta=1$). For $\delta \ll 1$, and $\gamma$ satisfying the stability requirement given after Eq.~\ref{KStraley} defects of class $C_w$ are
energetically favorable compared to those of the other classes, and
thus only these former defects survive until late times. For larger
values of $\delta$, but still less than $2\gamma$ (with $\gamma>0$) we
found defects belonging only to classes $C_u$ and $C_w$, consistent
with Eqs. (\ref{KStraley}), namely, $K_v>K_u+K_w$, and the results of
Ref. \cite{Zapotocky:95}. This regime also includes the
parameterization of Eq.~(\ref{Q}) used in
Refs.~\cite{Luckhurst:80,Biscarini:95}, where $\delta=4 \gamma^2/3$
was chosen. In Fig.~\ref{fig1} we show the total line length of defects in each of three classes as a function of time after the quench. In Fig.~\ref{fig2} we show a snapshot of the simulation cell late in the coarsening sequence. The $C_v$ class defects have disappeared and the $C_u$ and $C_w$ class defects remain as nonintersecting closed loops. When $\gamma<0$ and $\delta <2 |\gamma|$, we
have $K_u>K_v+K_w$, and we found, consistent with this inequality,
only defects of classes  $C_v$ and $C_w$ at late times. We found no
apparent line crossings, entanglements or junction points (the latter
obviously would require three classes of defects). The two classes of
loops appear to coarsen independently, as can be seen in the animations on our web site \cite{movies}.

\begin{figure}
\includegraphics[scale=0.5]{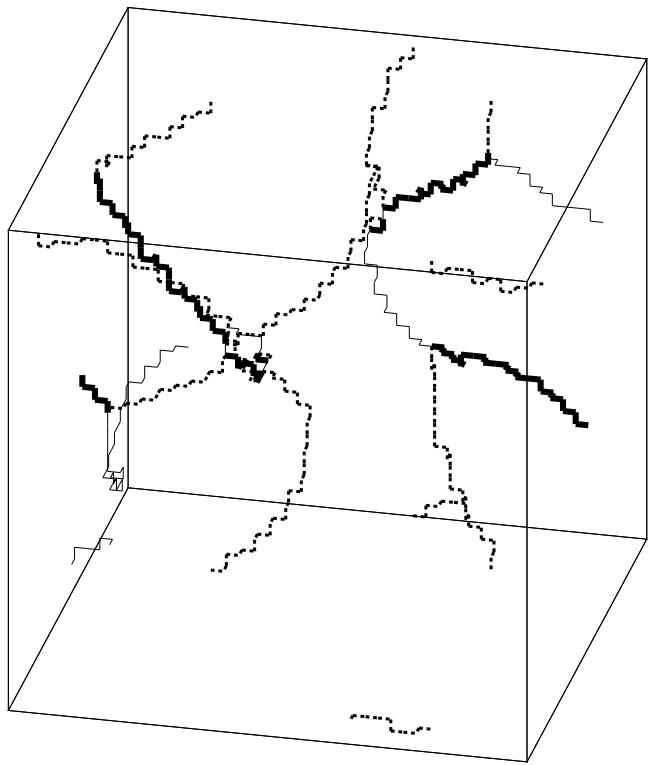}

\caption{The configuration of $C_{u}$ (thin line), $C_{v}$ (dotted line), and $C_{w}$ (bold line), 
 class defects for system size $40^{3}$, and parameters $\gamma=0$, $\delta=1$ ,  at dimensionless time 60 after the quench.}
\label{fig3}
%was fig 1
\end{figure}

\begin{figure}
\includegraphics[scale=0.5,angle=270]{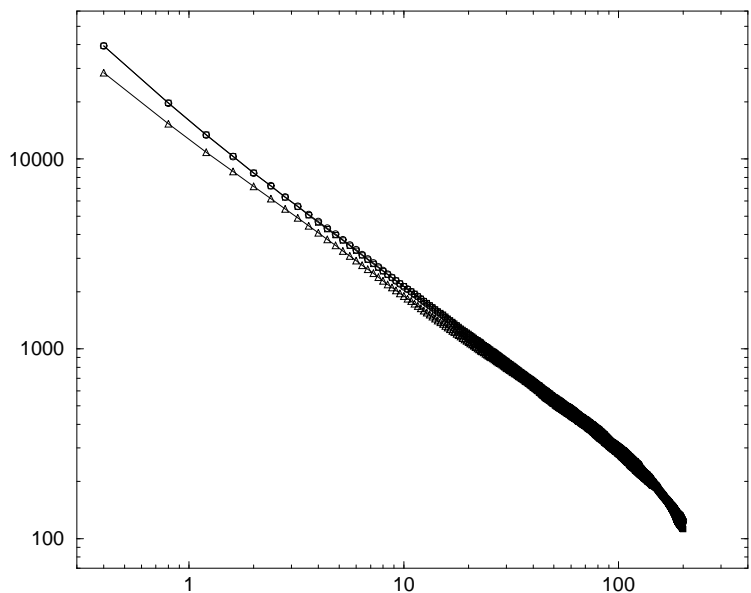}

\caption{Log-log plot of the total line length of $C_{v} \ (\circ), C_{w} \
(\rm{box})$, and $C_{u} \ (\triangle)$ class defects for system size $60^{3}$, and parameters 
$\gamma=0$, $\delta = 1$, as a function of  dimensionless time 
after the quench. The data has been averaged over 40 initial random configurations. 
While in principle the population of the three classes of defects should be equal immediately after the quench, there is a small difference in these populations due to the numerical issues discussed in the text following Eq.~(\protect\ref{D}).}
\label{fig4}
\end{figure}

The most interesting and novel coarsening sequence occurs when
$\gamma=0$ and $\delta=1$. For these parameters Eq.~(\ref{KStraley})
implies that all three elastic constants are equal, and we found a
coarsening sequence with nearly equal populations of the three
defects.  This behavior persists over a range of parameters $\delta \sim 1
\pm  0.5$ and $\gamma \sim 0\pm 0.2$. 
The coarsening sequence for these parameter values is particularly interesting in three dimensions. Soon after the quench a uniform network of junction points where three disclination lines (one from each of the three half--integer classes) meet is formed. These junction points are illustrated in Fig.~\ref{fig3} at late times after the quench. 
The junction points are distributed in a nearly uniform fashion throughout the simulation cell, and the
distance between neighboring points grows on average with time \cite{movies}. The total line length of each of the
three classes are nearly equal throughout the coarsening process (see Fig.~\ref{fig4}).
When the distance between neighboring junction points becomes comparable with the size
of the simulation cell (see Fig.~\ref{fig3}), the coarsening process is impeded. The final annihilation of the disclination lines can only occur via the shrinkage of individual loops. The formation of loops requires that some pairs of neighboring junction points approach each other, shrinking the line joining them while possibly increasing the length
of the other two disclination lines attached to the pair of junction points.
Ultimately, the pair of junction points meet at a ``pinch point'',
where four disclination line segments corresponding to two defect
classes meet. Subsequently the four line segments dissociate into two
nonintersecting single class line segments as can be seen in our animations \cite{movies}. When this process has occured a sufficient number of
times, individual disclination loops are formed which then  shrink independently.

In conclusion, we have shown that the coarsening dynamics of biaxial
nematics is very rich, with late time behavior governed by either one, two or three classes of half--integer line defects, depending on the parameters of the system. When all three classes are present, a network of junction points is formed and a novel coarsening sequence occurs.

We thank  S.~C. Ying, J.~M. Kosterlitz and M. Zapotocky for helpful discussions,
and G.~B. Loriot for computational assistance.
This work was supported by the National Science Foundation under grant DMR--9873849.
Computational work in support of this research was performed 
at Brown University's Theoretical Physics Computing Facility.

\begin{thebibliography}{17}
\expandafter\ifx\csname natexlab\endcsname\relax\def\natexlab#1{#1}\fi
\expandafter\ifx\csname bibnamefont\endcsname\relax
  \def\bibnamefont#1{#1}\fi
\expandafter\ifx\csname bibfnamefont\endcsname\relax
  \def\bibfnamefont#1{#1}\fi
\expandafter\ifx\csname citenamefont\endcsname\relax
  \def\citenamefont#1{#1}\fi
\expandafter\ifx\csname url\endcsname\relax
  \def\url#1{\texttt{#1}}\fi
\expandafter\ifx\csname urlprefix\endcsname\relax\def\urlprefix{URL }\fi
\providecommand{\bibinfo}[2]{#2}
\providecommand{\eprint}[2][]{\url{#2}}

\bibitem[{\citenamefont{Blundell and Bray}(1992)}]{Blundell:92}
\bibinfo{author}{\bibfnamefont{R.~E.} \bibnamefont{Blundell}} \bibnamefont{and}
  \bibinfo{author}{\bibfnamefont{A.~J.} \bibnamefont{Bray}},
  \bibinfo{journal}{Phys.~Rev.~A} \textbf{\bibinfo{volume}{46}},
  \bibinfo{pages}{R6154} (\bibinfo{year}{1992}).

\bibitem[{\citenamefont{Zapotocky et~al.}(1995)\citenamefont{Zapotocky,
  Goldbart, and Goldenfeld}}]{Zapotocky:95}
\bibinfo{author}{\bibfnamefont{M.}~\bibnamefont{Zapotocky}},
  \bibinfo{author}{\bibfnamefont{P.~M.} \bibnamefont{Goldbart}},
  \bibnamefont{and}
  \bibinfo{author}{\bibfnamefont{N.}~\bibnamefont{Goldenfeld}},
  \bibinfo{journal}{Phys.~Rev.~E} \textbf{\bibinfo{volume}{51}},
  \bibinfo{pages}{1216} (\bibinfo{year}{1995}).

\bibitem[{\citenamefont{Chuang et~al.}(1993)\citenamefont{Chuang, Yurke,
  Pargellis, and Turok}}]{Chuang:93}
\bibinfo{author}{\bibfnamefont{I.}~\bibnamefont{Chuang}},
  \bibinfo{author}{\bibfnamefont{B.}~\bibnamefont{Yurke}},
  \bibinfo{author}{\bibfnamefont{A.~N.} \bibnamefont{Pargellis}},
  \bibnamefont{and} \bibinfo{author}{\bibfnamefont{N.}~\bibnamefont{Turok}},
  \bibinfo{journal}{Phys.~Rev.~E} \textbf{\bibinfo{volume}{47}},
  \bibinfo{pages}{3343} (\bibinfo{year}{1993}).

\bibitem[{\citenamefont{Billeter et~al.}(1999)\citenamefont{Billeter,
  Smondyrev, Loriot, and Pelcovits}}]{Billeter:99}
\bibinfo{author}{\bibfnamefont{J.}~\bibnamefont{Billeter}},
  \bibinfo{author}{\bibfnamefont{A.~M.} \bibnamefont{Smondyrev}},
  \bibinfo{author}{\bibfnamefont{G.~B.} \bibnamefont{Loriot}},
  \bibnamefont{and} \bibinfo{author}{\bibfnamefont{R.~A.}
  \bibnamefont{Pelcovits}}, \bibinfo{journal}{Phys.~Rev.~E}
  \textbf{\bibinfo{volume}{60}}, \bibinfo{pages}{6831} (\bibinfo{year}{1999}).

\bibitem[{\citenamefont{Toulouse}(1977)}]{Toulouse:77}
\bibinfo{author}{\bibfnamefont{G.}~\bibnamefont{Toulouse}},
  \bibinfo{journal}{J. Phys. Lett. (Paris)} \textbf{\bibinfo{volume}{38}},
  \bibinfo{pages}{L67} (\bibinfo{year}{1977}).

\bibitem[{\citenamefont{Mermin}(1979)}]{Mermin:79}
\bibinfo{author}{\bibfnamefont{N.~D.} \bibnamefont{Mermin}},
  \bibinfo{journal}{Rev.~Mod.~Phys.} \textbf{\bibinfo{volume}{51}},
  \bibinfo{pages}{591} (\bibinfo{year}{1979}).

\bibitem[{\citenamefont{De'Neve et~al.}(1992)\citenamefont{De'Neve, Kleman, and
  Navard}}]{Deneve:92}
\bibinfo{author}{\bibfnamefont{T.}~\bibnamefont{De'Neve}},
  \bibinfo{author}{\bibfnamefont{M.}~\bibnamefont{Kleman}}, \bibnamefont{and}
  \bibinfo{author}{\bibfnamefont{P.}~\bibnamefont{Navard}},
  \bibinfo{journal}{J. Phys. II (France)} \textbf{\bibinfo{volume}{2}},
  \bibinfo{pages}{187} (\bibinfo{year}{1992}).

\bibitem[{\citenamefont{Poenaru and Toulouse}(1977)}]{Poenaru:77}
\bibinfo{author}{\bibfnamefont{V.}~\bibnamefont{Poenaru}} \bibnamefont{and}
  \bibinfo{author}{\bibfnamefont{G.}~\bibnamefont{Toulouse}},
  \bibinfo{journal}{J. Phys. (Paris)} \textbf{\bibinfo{volume}{8}},
  \bibinfo{pages}{887} (\bibinfo{year}{1977}).

\bibitem[{\citenamefont{Kobdaj and Thomas}(1994)}]{Kobdaj:94}
\bibinfo{author}{\bibfnamefont{C.}~\bibnamefont{Kobdaj}} \bibnamefont{and}
  \bibinfo{author}{\bibfnamefont{S.}~\bibnamefont{Thomas}},
  \bibinfo{journal}{Nucl. Phys. B} \textbf{\bibinfo{volume}{413}},
  \bibinfo{pages}{689} (\bibinfo{year}{1994}).

\bibitem[{\citenamefont{Sukumaran and Ranganath}(1997)}]{Sukumaran:97}
\bibinfo{author}{\bibfnamefont{S.}~\bibnamefont{Sukumaran}} \bibnamefont{and}
  \bibinfo{author}{\bibfnamefont{G.~S.} \bibnamefont{Ranganath}},
  \bibinfo{journal}{J. Phys. II (France)} \textbf{\bibinfo{volume}{7}},
  \bibinfo{pages}{583} (\bibinfo{year}{1997}).

\bibitem[{\citenamefont{Straley}(1974)}]{Straley:74}
\bibinfo{author}{\bibfnamefont{J.~P.} \bibnamefont{Straley}},
  \bibinfo{journal}{Phys. Rev. A} \textbf{\bibinfo{volume}{10}},
  \bibinfo{pages}{1881} (\bibinfo{year}{1974}).

\bibitem[{\citenamefont{Luckhurst and Romano}(1980)}]{Luckhurst:80}
\bibinfo{author}{\bibfnamefont{G.~R.} \bibnamefont{Luckhurst}}
  \bibnamefont{and} \bibinfo{author}{\bibfnamefont{S.}~\bibnamefont{Romano}},
  \bibinfo{journal}{Mol. Phys.} \textbf{\bibinfo{volume}{40}},
  \bibinfo{pages}{129} (\bibinfo{year}{1980}).

\bibitem[{\citenamefont{Biscarini et~al.}(1995)\citenamefont{Biscarini,
  Chiccoli, Pasini, Semeria, and Zannoni}}]{Biscarini:95}
\bibinfo{author}{\bibfnamefont{F.}~\bibnamefont{Biscarini}},
  \bibinfo{author}{\bibfnamefont{C.}~\bibnamefont{Chiccoli}},
  \bibinfo{author}{\bibfnamefont{P.}~\bibnamefont{Pasini}},
  \bibinfo{author}{\bibfnamefont{F.}~\bibnamefont{Semeria}}, \bibnamefont{and}
  \bibinfo{author}{\bibfnamefont{C.}~\bibnamefont{Zannoni}},
  \bibinfo{journal}{Phys. Rev. Lett.} \textbf{\bibinfo{volume}{75}},
  \bibinfo{pages}{1803} (\bibinfo{year}{1995}).

\bibitem[{\citenamefont{Bac et~al.}(2001)\citenamefont{Bac, Peredez, Vasquez,
  Medina, and Hasmy}}]{Bac:01}
\bibinfo{author}{\bibfnamefont{C.~G.} \bibnamefont{Bac}},
  \bibinfo{author}{\bibfnamefont{R.}~\bibnamefont{Peredez}},
  \bibinfo{author}{\bibfnamefont{C.}~\bibnamefont{Vasquez}},
  \bibinfo{author}{\bibfnamefont{E.}~\bibnamefont{Medina}}, \bibnamefont{and}
  \bibinfo{author}{\bibfnamefont{A.}~\bibnamefont{Hasmy}},
  \bibinfo{journal}{Phys. Rev. E} \textbf{\bibinfo{volume}{63}},
  \bibinfo{pages}{042701} (\bibinfo{year}{2001}).

\bibitem[{\citenamefont{Lammert et~al.}(1995)\citenamefont{Lammert, Rokhsar,
  and Toner}}]{Lammert:95}
\bibinfo{author}{\bibfnamefont{P.~E.} \bibnamefont{Lammert}},
  \bibinfo{author}{\bibfnamefont{D.~S.} \bibnamefont{Rokhsar}},
  \bibnamefont{and} \bibinfo{author}{\bibfnamefont{J.}~\bibnamefont{Toner}},
  \bibinfo{journal}{Phys.~Rev.~E} \textbf{\bibinfo{volume}{52}},
  \bibinfo{pages}{1778} (\bibinfo{year}{1995}).

\bibitem[{\citenamefont{Priezjev and Pelcovits}(2001)}]{Priezjev:01}
\bibinfo{author}{\bibfnamefont{N.~V.} \bibnamefont{Priezjev}} \bibnamefont{and}
  \bibinfo{author}{\bibfnamefont{R.~A.} \bibnamefont{Pelcovits}},
  \bibinfo{journal}{Phys. Rev. E} \textbf{\bibinfo{volume}{64}},
  \bibinfo{pages}{031710} (\bibinfo{year}{2001}).

\bibitem[{mov()}]{movies}
\eprint{http://www.physics.brown.edu/Users/faculty/pelcovits/biaxial/}.

\end{thebibliography}
\end{document}